\documentclass[pra,twocolumn,aps,amssymb,footinbib,showpacs]{revtex4-1}
\usepackage{amssymb}
\usepackage{amsmath}
\usepackage{times}
\usepackage{latexsym}
\usepackage{graphicx}
\usepackage{color}
\usepackage{bm}
\usepackage{subfigure}
\newcommand{\br}{\mathbf{r}}

\begin{document}
\title{Cosine Edge Mode in a Periodically Driven Quantum System}
\author{ Indubala I. Satija and Erhai Zhao}
\affiliation{
Department of Physics and Astronomy, George Mason University, Fairfax, VA 22030, USA}
\begin{abstract}
{ 
Time-periodic (Floquet) topological phases of matter exhibit bulk-edge relationships that are more complex 
than static topological insulators and superconductors. Finding the edge modes unique to driven systems usually requires 
numerics. Here we present a minimal two-band model of Floquet topological insulators and semimetals in two dimensions 
where all the bulk and edge properties can be obtained analytically. 
It is based on the extended Harper model of quantum Hall effect at flux one half. 
We show that periodical driving gives rise to a series of phases characterized by a pair of integers. 
The model has a most striking feature: the spectrum of the edge modes is always given by a single cosine function, $\omega(k_y)\propto \cos k_y$ where
$k_y$ is the wave number along the edge, as if it is freely dispersing and completely decoupled from the bulk.
The cosine mode is robust against the change in driving parameters and persists even to semi-metallic phases with Dirac points.
The localization length of the cosine mode is found to contain an integer and in this sense quantized.
}
\end{abstract}

\maketitle

Robust boundary or edge modes are hallmarks of topological insulators and superconductors \cite{RevModPhys.82.3045,RevModPhys.83.1057}. They can be viewed as the ``holographic duals" of the bulk through the bulk-boundary correspondence \cite{RevModPhys.83.1057}. While the existence of the edge modes are intuitively understood by the standard topological arguments and firmly established mathematically, e.g., by the index theorem, finding their exact dispersion from the bulk Hamiltonian usually requires involved procedures. Take the Harper model of integer quantum Hall effect for example \cite{harper,hofstadter_energy_1976}. It describes non-interacting fermions on two dimensional (2D) lattices in the presence of a magnetic field. A standard way to obtain its edge mode is to introduce the transfer matrix and then solve a higher order equation \cite{hatsugai,bernevig2013topological}. Alternatively, the Hamiltonian of a finite system, such as a slab, is diagonalized numerically. The dispersions of the chiral edge modes are rarely given by simple, analytic functions. 

Recently, time-periodic quantum systems, such as a piece of graphene irradiated by a driving light field \cite{oka_2009}, are found to develop interesting topological phases and edge modes that may or may not have a static analog. New concepts are introduced to describe the unique properties of these so-called Floquet topological insulators \cite{oka_2009,lindner_floquet_2011}. A number of topological invariants have been constructed from the time-evolution operator $U(t)$ \cite{kitagawa_topological_2010, rudner_anomalous_2013, PhysRevLett.114.106806}. For 2D lattice systems, the point-like singularities in the phase bands during the time evolution are related to the winding number which is equal to the net chirality of the edge modes in given quasienergy band gap \cite{nathan}. Non-interacting Floquet topological insulators can be classified according to the Altland-Zirnbauer symmetry classes and spatial dimensions by decomposing the unitary evolution into two parts \cite{roy2016periodic}. For example, a 2D Floquet insulator in class A is characterized by $Z\times Z$, i.e. a pair of integers, rather than the familiar $Z$ or $Z_2$ number. Despite the progress, the bulk-boundary correspondence in Floquet systems remains only partially understood \cite{roy2016periodic}. Compared to their static counter-parts, the Floquet edge modes are intrinsically more complex. Finding their dispersion relies even more on numerical analysis. It is therefore desirable to construct models for which the Floquet edge modes are described by elementary functions.

In this letter, we present an analytically solvable periodically driven lattice model in two dimensions. It only has two bands and generalizes the Harper model at flux one-half ($\pi$ flux) by allowing the hopping amplitudes to vary periodically in time. The time evolution of this Floquet system takes the form of (momentum-dependent) successive rotations of pseudo-spin $1/2$. This makes it possible to analyze its topological properties analytically. Although the phase diagram of the model contains a rich collection of topologically distinct phases characterized by the Chern and winding numbers, the Floquet edge states are invariably described by a single cosine function across all quasienergy gaps and for all insulating and semi-metallic phases. Relatedly, the localization (decay) length of the cosine edge mode is dictated by an analytical formula that resembles a quantization condition. 

{\it Extended Harper model}. Consider spinless fermions hopping on a square lattice within the $xy$ plane subject to a magnetic field in the $z$-direction \cite{hofstadter_energy_1976}. The magnetic flux threading each square plaquette is set to $\phi$ in unit of the flux quantum. 
Following Thouless \cite{Thou83}, we further include hopping $J_d$ between the next-nearest-neighbor sites (along the diagonals of the square lattice).
The resulting system, referred to as the {extended Harper model}, is described by the Hamiltonian
\begin{eqnarray}
H_s  &=& -\sum_{\br} [J_x c_{\br+\hat{x}}^{\dagger} c_{\br} + J_y e^{i2 n\pi \phi }c_{\br+\hat{y}}^{\dagger} c_{\br} \nonumber \\
&+&J_d  e^{i (n+\frac{1}{2})2 \pi \phi }(c_{\br+\hat{x}+\hat{y}}^{\dagger}+c_{\br-\hat{x}+\hat{y}}^{\dagger}) c_{\br}] +h.c.
\label{Hamil}
\end{eqnarray}
where $\br=n\hat{x}+m\hat{y}$ labels the lattice sites with $n,m$ being integers, and $c^\dagger_\br$ creates a fermion at site $\br$. We set the lattice spacing to be one and work in the Landau gauge, so the vector potential $A_x=0$ and $A_y = n \phi h/e$. 
The model for arbitrary flux $\phi$ was discussed in detail in Refs. \onlinecite{Hat,Thou94}. A particularly fascinating aspect of the system is a new type of critical phase termed ``bicritical" when ${J_d}/{J_x}$ exceeds  $1/2$ \cite{Thou94}.
Recently,  its highly nontrivial mathematical properties  were analyzed  in Ref. \onlinecite{Gharper}. 

We will focus on the case $\phi=1/2$ as recently realized in cold atoms experiment \cite{mit-exp}. Then $H_s$ reduces to a two-band model. It is well known that if only nearest neighbor hoppings $J_{x}$ and $J_y$ are allowed, its spectrum is gapless and Dirac-like around zero energy. For finite diagonal hopping, $\alpha \equiv  {J_d}/{J_x}\neq 0$, the energy spectrum $E_k=\pm2J_x(\cos^2k_x+ \lambda \cos^2 k_y+4\alpha^2 \sin^2k_x\sin^2k_y)^{1/2}$ becomes gapped
with Chern numbers of the two bands being $\pm 1$ \cite{Hat}. Here $\lambda=J_y/J_x$ is the $x-y$ hopping anisotropy. 

{\it  A two-band model of Floquet topological matter}. We now generalize $H_s$ into a periodically driven model by allowing the hopping amplitudes to vary periodically with time $t$, following Ref. \onlinecite{LSZ,ZhouPRB}. Assume that for $0<t<T_1$, only the $x$-hopping $J_x$ and the diagonal hopping $J_d$ are present. In crystal momentum space, the Hamiltonian is a $2\times 2$ matrix, 
\begin{equation}
  H_1  =    -2J_x\cos k_y \sigma_x +4J_d\sin k_x \sin k_y\sigma_y. 
\end{equation}
Here the $\sigma$'s are Pauli matrices in the pseudo-spin space describing the sublattice degrees of freedom (each unit cell contains two sites). 
For $T_1<t<T$, i.e. for a duration of $T_2\equiv T-T_1$, only $J_y$ is turned on,
\begin{equation}
  H_2  =    -2J_y\cos k_y \sigma_z.
\end{equation}
The Hamiltonian is periodic in time, $H(t+T)=H(t)$, and piecewise constant. We will refer to this as square wave driving \cite{LSZ}. The time evolution operator
for a full period $T$ consists of two successive rotations  \cite{zhao2016anatomy} in spin space, 
\begin{equation}
U(T) = e ^{- i H_2 T_2} e ^ {-i H_1 T_1}=e^{i \chi_2  \sigma_z} e^{i \chi_1  (\cos \zeta\sigma_x + \sin \zeta\sigma_y)}, \label{uop}
\end{equation}
with the two rotation angles given by 
\begin{eqnarray}
 \chi_1  &=&  2J_x T_1 \sqrt{ \cos ^2 k_x + 4 \alpha^2 \sin^2 k_x \sin^2 k_y}, \nonumber \\
 \chi_2  &=& 2J_y T_2 \cos k_y, \label{ang}
\end{eqnarray}
and $\tan\zeta=2 \alpha \tan k_x \sin k_y$. 

Alternatively, we can generalize $H_s$ to a periodically kicked model \cite{LSZ,PhysRevB.90.195419}. Assume 
$J_x$ and $J_d$ are
held constant, while $J_y$ is only turned on when $t$ is multiples of the period $T$,
\begin{equation}
{H}(t) = {H_1}+{H_2} \sum_m \delta(t/T-m),
 \label{ht}
\end{equation}
where $m$ is an integer. The one-kick evolution operator $U(T)$ is still given by Eq. \eqref{uop}-\eqref{ang} with the replacement $T_1, T_2\rightarrow T$.
In fact, periodical kicking can be viewed as the following limit of square wave driving: $T_1\rightarrow T$, $T_2\rightarrow 0$ with $J_yT_2$ fixed at some constant. The topological properties of these two types of models are thus identical. Without loss of generality, we will focus on the kicked model below.  Its parameter space includes the hopping ratio $\alpha$ and two dimensionless driving parameters $\bar{J_x} \equiv J_xT/\pi$ and $\bar{J_y}  \equiv J_yT/\pi$. 
 
The eigenvalues of $U(T)$ have the form $e^{-i\omega_n T}$ with $\omega_n$ called the quasienergy. The effective Hamiltonian is defined by 
 $U(T)= e^{- i H_\mathrm{eff} T}$. Even though $H_1$ and $H_2$ do not commute, the two rotations in $U(T)$ can be combined into a single rotation around some axis $\hat{n}$ by an angle $\omega T$,
 \begin{equation}
U(T) =  e^{ i \omega T \boldsymbol{\sigma} \cdot \hat{n} }.
\label{U2}
\end{equation}
Eq. \eqref{U2} automatically diagonalizes $U$. The quasienergies are just $\pm\omega$, reflecting particle-hole symmetry, with $\omega$ given by 
\begin{equation}
\cos (\omega T) = \cos \chi_2 \cos \chi_1.
\label {w}
\end{equation}
Eq. \eqref{w} is one of the key results of our paper. The effective Hamiltonian has the form of quantum spin 1/2 in a $\mathbf{k}$-dependent 
magnetic field, $H_\mathrm{eff}(\mathbf{k}) = \boldsymbol{\sigma} \cdot \mathbf{B}_\mathrm{eff}(\mathbf{k})$ with $\mathbf{B}_\mathrm{eff}(\mathbf{k})=\omega T \hat{n}(\mathbf{k})$. The direction $\hat{n}(\mathbf{k})$ is given by $n_x=\sin \chi_1\cos(\chi_2+\zeta)/\sin(\omega T)$, $n_y=\sin \chi_1\sin(\chi_2+\zeta)/\sin(\omega T)$, and $n_z=\cos \chi_1\sin\chi_2/\sin(\omega T)$.

\begin{figure}[htbp]
\includegraphics[width = 0.7 \linewidth]{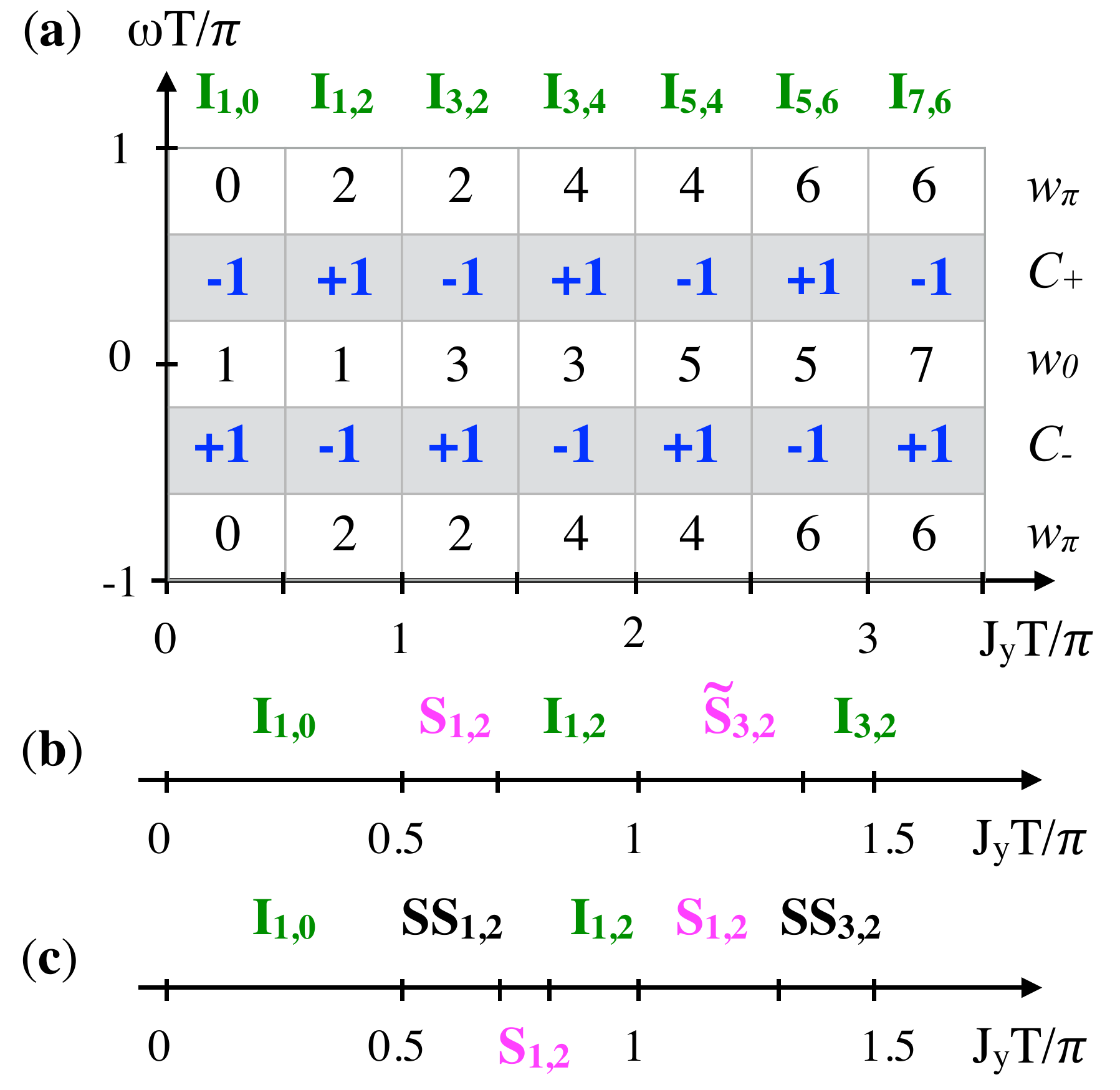}
\caption{
Phase diagrams of the periodically kicked model Eq. \eqref{ht}. 
(a) Schematic of the quasienergy bands (shaded region) and gaps (empty region) as $\bar{J_y}$ is increased for fixed $\bar{J_x}<1/2$ and $\alpha<1/2$. A series of insulating (I) phases I$_{w_0,w_{\pi}}$ are identified by examining the band Chern numbers $C_+$, $C_-$ and the winding numbers $w_0$, $w_{\pi}$.  
(b) Topological phases with fixed $\alpha=1/2$ and $\bar{J_x}=0.75$. S$_{1,2}$ and $\tilde{\mathrm{S}}_{3,2}$ 
are semi-metallic (S) phases described in the main text. Their spectra are shown in Fig. \ref{f2}(b) and \ref{f2}(d). 
(c) Topological phases along the line $2\alpha=\bar{J_x}=\bar{J_y}$. SS denotes (double) semi-metallic phases with Dirac points at both quasienergy 0 and $\pi/T$.
%
}
\label{f1}
\end{figure}

\begin{figure}[htbp]
\includegraphics[width = 0.9\linewidth]{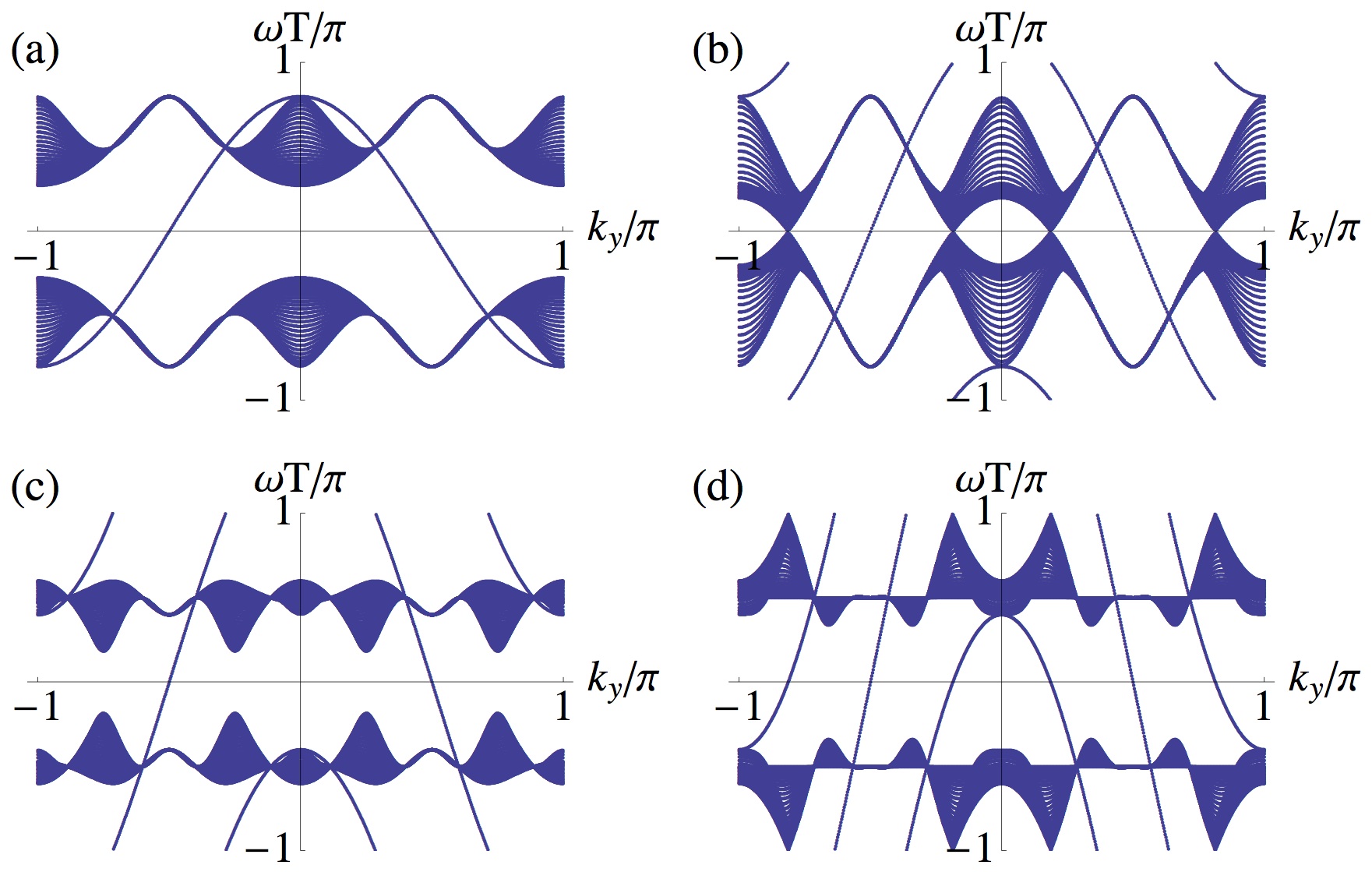}
\caption{ Quasienergy spectrum of a slab of finite width $L_x=61$ in the $x$-direction but periodic in $y$-direction. Panel (a), (b),
(c), (d) are for $\bar{J_y}=0.4$, $0.6$, $0.8$ and $1.2$, corresponding to phase I$_{1,0}$, S$_{1,2}$,  I$_{1,2}$, and $\tilde{\mathrm{S}}_{3,2}$. Here $\alpha=1/2$, $\bar{J}_x=\bar{J}_y$ except for (d), $\bar{J_x}=0.75$. 
}
\label{f2}
\end{figure}

The periodically kicked model Eq. \eqref{ht} features a rich collection of (Floquet) phases as the parameters $\alpha$,  $\bar{J_x}$ and $\bar{J_y}$ are varied. Each phase has its own characteristic bulk quasienergy spectrum and the associated topological invariants and edge states. The phase diagram can be determined by analyzing Eq. (\ref{w}). Two examples along different cuts in the parameter space are given in Fig. \ref{f1}. Fig. \ref{f2} illustrates the quasienergy spectra of four representative phases in the slab geometry where both the bulk band structure and the edge states are visible.

{\it  Insulating phases characterized by a pair of integers}.  The phase diagram Fig. \ref{f1}(a) for $\bar{J}_x<1/2$ and fixed $\alpha<1/2$ is very simple. A series of phases, labeled by I$_{1,0}$, I$_{1,2}$ etc., appear consecutively as $\bar{J}_y$ goes through integer multiples of $1/2$. 
All these phases have two finite gaps at quasienergy $0$ and $\pi/T$ and two well separated bands. The 0-gap and $\pi$-gap are characterized by the winding number $w_0$ and $w_{\pi}$ respectively. And we denote the Chern number for the band at positive (negative) energies by $C_+$ ($C_-$). The Chern numbers and the winding numbers are related by, e.g., $C_{+}=w_{\pi}-w_0$ and $C_+=-C_+$. Thus, each  phase can be labelled by a pair of integers $w_0$ and $w_{\pi}$ (or equivalently $C_+$ and $w_0$ etc.). It is a Floquet Insulator (I), so we refer to it as phase I$_{w_0,w_{\pi}}$. In particular, phase I$_{1,0}$ corresponding to fast driving (small $T$) is identical to the static quantum Hall state at flux 1/2 analyzed in Ref. \onlinecite{Hat}. As shown in Fig. \ref{f2}(a), $w_\pi=0$ implies no edge states inside the $\pi$-gap. In contrast, all the other Floquet phases in Fig. \ref{f1}(a) have finite number of chiral edge modes inside the $\pi$-gap [e.g. I$_{1,2}$ in Fig. \ref{f2}(c)] and therefore they {\it have no static analog}. 
 
One notices that in Fig. 1(a), the Chern number $C_+$ simply alternates between 1 and -1, and the phase transition points are equally distributed. These can be understood from Eq. (\ref{w}). For small $\bar{J}_x$ and $\chi_1$, $\cos (\omega T) \simeq \cos \chi_2$. So the quasienergy $\omega$ crosses $0$ or $\pi/T$ when $\chi_2$ equals to $2n\pi$ or $(2n+1)\pi$ for integer $n$. In either cases, the two quasienergy bands touch each other, triggering a change in the band Chern number. According to Eq. \eqref{ang}, this occurs at $J_yT=n\pi$ where the gap closes at zero energy and $w_0$ changes by 2, or at $J_yT=(n+1/2)\pi$ where the $\pi$-gap closes and $w_\pi$ jumps by 2. 

The sequence of odd $w_0$ appearing at the 0-gap and even $w_{\pi}$ at the $\pi$-gap found here in Fig. \ref{f1}(a) 
is reminiscent of a similar sequence in the static extended Harper model in the vicinity of flux $1/2$ as described in Ref. \cite{Satijabook}. 
It seems as if by varying the parameter $\bar{J}_y$,  the driven system for fixed flux is able to ``access" the topological edge states of various gaps of the corresponding static system with a flux value slightly away from $1/2$. We recall that the winding numbers correspond to the (infinitely many) solutions of the Diophantine equation~\cite{Dana85}. As a result, the entries in each sequence are related by modulo $2$, the denominator of the flux $1/2$.

{\it Semi-metallic phases and Dirac points}.
The phase diagram becomes more complicated for larger values of $\bar{J_x}$ and $\alpha$.
One example is shown in Fig. 1(b) along for fixed $\bar{J}_x=0.75$ and $\alpha=1/2$. Besides the insulating phases with two gaps such as I$_{1,0}$ and I$_{1,2}$ discussed above, new phases emerge which have only {\it one} well-defined quasienergy gap but Dirac points at $\omega=0$ or $\pi/T$ in the spectrum. We refer to them as semi-metallic (S) phases. For example, the phase S$_{1,2}$ is gapped at $\omega=\pi/T$ but gapless at $\omega=0$. Its spectrum shown in Fig. \ref{f2}(b) is analogous to the familiar Dirac semimetal: the two bands become degenerate at zero energy for certain $\mathbf{k}$-points. These locations can be found from Eq. (\ref{w}). A solution for $\omega=0$ requires that $ \cos \chi_2 \cos \chi_1=1$, i.e., 
\begin{equation}
\chi_{1}   = n_1 \pi, \quad
\chi_{2} =  n_2 \pi, \quad n_1+n_2=even,
\label{DD2}
\end{equation}
which only has solutions at isolated $\mathbf{k}$-points according to Eq. \eqref{ang}. The spectrum of the other semi-metallic phase $\tilde{\mathrm{S}}_{3,2}$ is illustrated in Fig. \ref{f2}(d). In contrast to S$_{1,2}$, it is gapped at $\omega=0$ but has Dirac points at quasienergy $\omega=\pi/T$. It is therefore a Floquet semimetal \cite{PhysRevE.93.022209}. We distinguish it from S by tilde to emphasize that it has no counterpart in the static model $H_s$. The locations of its Dirac points are also given by Eq. \eqref{DD2} but with $n_1+n_2=odd$.

Floquet semimetals have yet to be classified systematically and there is no generally accepted convention to label them. Here we find it proper to identify the semi-metallic phases unambiguously using a pair of integers $(w_0,w_\pi)$, e.g. S$_{w_0,w_\pi}$ and $\tilde{\mathrm{S}}_{w_0,w_\pi}$, to indicate the number of edge modes at quasienergy 0 and $\pi/T$, as done above for the insulating states. It is attempting to label the phase S$_{1,2}$ by only one integer $w_{\pi}=2$. But this is insufficient and misses one important point: as shown in Fig. \ref{f2}(b),  edge modes are present not only inside the $\pi$ gap but also {\it at all other energies coexisting with the bulk states}, including the Dirac points at $\omega=0$. In the latter case, the edge states appear to be ``caged", i.e. bounded in $k_y$ by the Dirac points. They are robust and can be viewed as the continuation of the edge states of phase I$_{1,2}$ despite the gap closing at the Dirac points. 
Similar coexistence of the edge and bulk states is also observed in phase $\tilde{\mathrm{S}}_{3,2}$ and many other semi-metallic phases, such as
the SS phases with Dirac points both at quasienergy zero and $\pi/T$ shown in Fig. \ref{f1}(c).
To summarize, the semi-metallic states found here are distinct from static Dirac semimetals or gapless Floquet superconductors studied in Ref. \cite{zhao2016anatomy}. They are characterized by two integers.
%
The persistence of the edge modes into the semi-metallic phases will become more transparent once we work out its analytical expression below.

{\it Cosine edge mode and its localization length}.
Given the plethora of insulating and semi-metallic phases found in this model, it is natural to expect that the energy-momentum dispersion of the edge states,  $\omega_\mathrm{edge}(k_y)$, to change from one phase to another or from one gap to another as in the static Hofstadter problem. It thus comes as a surprise that $\omega_\mathrm{edge}(k_y)$ takes the same simple form for all phases, including the semi-metallic phases, regardless of the driving parameters. Empirically, the functional form of $\omega_\mathrm{edge}(k_y)$ becomes apparent if the spectra in Fig. 2 are plotted in the repeated zone scheme for both the quasienergy and the quasimomentum.

We solve for $\omega_\mathrm{edge}(k_y)$ from the bulk dispersion Eq. \eqref{w} and show it is simply the cosine function by using a well established method in band theory \cite{kohn,heine,PhysRevB.25.3975}. In order to enlarge the Hilbert space to accommodate localized states, we allow Bloch wavevectors $k$ to take complex values and analytically continue $H(k)$ (and $U$) which  becomes non-Hermitian in general.  The edge states correspond to {\it real} eigenenergies (inside the bulk gap) of $H(k)$ for complex values of $k$. For the slab geometry in our case, it is sufficient to let $k_x=k_r+i k_i$ in Eq. \eqref{w}, where $k_r$, $k_i \in \mathbb{R}$ are the real and imaginary part of $k_x$ respectively. In order to guarantee a real solution for $\omega$, we must require $\chi_1$ to be real, which in turn requires that, after a little algebra, $\sin (2 k_r)=0$. It has two types of solutions: 
\begin{eqnarray}
&\mathrm{I}: & k_r = \pm\pi/2; \label{type1} \\
&\mathrm{II}: & k_r = 0. \label{type2}
\end{eqnarray}

The numerical fact that the edge spectrum does not depend on $J_x$ for fixed $J_y$ implies that $\cos\chi_1$ collapses to a constant. The continuity of the edge spectrum in the limit of $J_xT\rightarrow 0$ further fixes the constant to be 1. Namely, $\cos\chi_1=1$ or
\begin{equation}
\chi_1 =2n_e\pi, \;\;\; n_e\in \mathbb{Z}.
\label{quantization}
\end{equation}
Then Eq. \eqref{w} simplifies to $\cos (\omega \tau) = \cos \chi_2$ which leads to
\begin{equation}
\omega_\mathrm{edge} (k_y) =  2 J_y \cos k_y.
\label{en}
\end{equation}
We have checked that Eq. \eqref{en} fits exactly the numerically obtained edge spectra of finite slabs, e.g., those in Fig. 2. The inverse localization/decay length $k_i$ of the edge mode can be found from Eq. \eqref{quantization} and the expression for $\chi_1$,
\begin{eqnarray}
 \sinh^2k^\mathrm{I}_i &=& \frac{1-( n_e/\bar{J}_x)^2}{ 1-(2\alpha \sin k_y)^2 } - 1, \\
 \sinh^2k^{\mathrm{II}}_i &= &\frac{(n_e/\bar{J}_x)^2 - 1}{ 1-(2\alpha \sin k_y)^2 },
\label{L1}
\end{eqnarray}
for type I and type II solution respectively.


The decay length appears to be ``quantized" due to the presence of integer $n_e$ in Eq. \eqref{L1}.  The natural question is:  {\it how does the system choose this quantum number}?  Our detailed numerical studies of the slab geometry reveals that $n_e$ is nothing but the integer part of $\bar{J}_x$,
\begin{equation}
n_e = [ \bar{J}_x ].
\label{ne}
\end{equation}
The value of $n_e$ seems to encode the condition that  the decay length  diverges when the cosine edge mode merges with the bulk bands.
Presently, a deeper understanding of this simple but fascinating result remains elusive.

In the fast driving limit, $\bar{J}_x \rightarrow 0$,  the localization lengths of the edge modes correspond to $n_e=0$ and therefore the static
 model permits only type I localization. In the driven model,
 for $\alpha < 1/2$, the edge states are of type I while for $\alpha > 1/2$, they switch between type I and type II as $k_y$ varies. Interestingly,
the switch from type I to II occurs at values of $k_y=k^*_y$ where the edge states cross bulk at special points where the bulk spectrum is ``pinched", i.e. independent of $k_x$. The pinching occurs when $2\bar{J_y} \cos k^*_y =  \pm  (2m+1)\pi/2$, corresponding to quasienergy  $\pm \pi/2$.  We note that the existence of two distinct types of localized edge modes in our model is reminiscent  of two types of localized regimes in the static extended Harper model for irrational flux~\cite{Thou94}.  However, understanding the possible relationships between the localization characteristics of the static and the driven system is still lacking.

{\it Discussions}. Our periodically kicked model thus presents a marked dichotomy: the dispersion of the cosine edge state only depends on $\bar{J_y}$, while its decay length only depends on $\bar{J_x}$. On the $\bar{J_x}-\bar{J_y}$ plane, for increasing $\bar{J_y}$ but fixed $J_x$, the cosine mode will stretch in amplitude and continuously wind cross the quasienergy Brillouin zone boundary $\pi/T$, giving rise to ever-increasing number of chiral edge modes in the 0 and $\pi$ gap shown in Fig. 1(a). When $\bar{J}_x$ is increased for fixed $\bar{J}_y$, e.g. $\bar{J}_y=1/4$, each time $\bar{J}_x$ reaches an integer value, $n_e$ jumps by one and the gap closes at $\omega=0$. The distinction between the bulk and edge states gets lost but 
%
the edge state spectrum stays the same.

The simple cosine dispersion of the edge mode in our Floquet system is identical to that of a free particle hopping on a one-dimensional chain along the $y$ direction. In other words, the edge states seem ``perfectly" localized as if they do not couple at all to sites away from the edge, even though the hopping amplitudes $J_x$ and $J_d$ are finite. The lack of diffusion is reminiscent of Creutz's ice-tray model \cite{RevModPhys.73.119}, where edge states form at the ends of a two-leg ladder with $J_x$ and $J_d$ at flux $1/2$ due to deconstructive interference. Note however $H_1$ in our model is periodic in $y$ and not one-dimensional. Furthermore, the wave function of the cosine edge mode varies with time.
One single function describing all the edge modes reveals the continuity of the edge spectrum and the robustness of the edge state throughout the phase diagram. For example, the persistence of the edge mode into the semi-metallic phases become very natural and easy to understand. Similar picture emerges for other driving protocols, e.g. with $J_d$ time-dependent but ${J_x}$, ${J_y}$ held constant, and also for triangular lattice.

In summary, the model introduced and solved here brings a new perspective to the active field of systematically understanding the topological properties of time-periodic quantum systems. Our simple Floquet system exhibits a rich variety of topological insulating phases I$_{w_0,w_\pi}$ characterized by a pair of integers as well as semi-metallic phases S$_{w_0,w_\pi}$, $\tilde{\mathrm{S}}_{w_0,w_\pi}$ with Dirac points developed at quasienergy zero or $\pi/T$ and Floquet edge states existing at all quasienergies. Its most remarkable feature is the simple cosine dispersion of the edge state across the entire phase diagram. The cosine edge mode also shows a nontrivial behavior in its decay length. In addition to these new features and the pedagogical value of the model itself, the construction may be generalized to analyze driven systems in other symmetry classes and spatial dimensions.

This work is supported by by AFOSR grant number FA9550-16-1-0006 and NSF PHY-1205504 (EZ).

\bibliography{cosIIS}

\end{document}